\PassOptionsToPackage{table}{xcolor}
\documentclass[letterpaper]{article}
\usepackage[preprint]{aaai2027}

\usepackage[hyphens]{url}
\usepackage{graphicx}
\usepackage{natbib}
\usepackage{caption}
\usepackage{amsmath}
\usepackage{amsthm}
\usepackage{amssymb}
\usepackage{mathtools}
\usepackage{enumitem}
\usepackage{booktabs}
\usepackage{makecell}
\usepackage{multirow}
\usepackage{tabularx}
\usepackage{threeparttable}
\usepackage{subcaption}
\usepackage[most]{tcolorbox}
\usepackage{algorithm}
\usepackage{algorithmic}

\definecolor{gapzero}{RGB}{46,125,50}
\definecolor{gapsmall}{RGB}{102,187,106}
\definecolor{gaplarge}{RGB}{102,187,106}
\definecolor{tprgood}{RGB}{56,142,60}
\definecolor{tprok}{RGB}{102,187,106}
\definecolor{na}{gray}{0.75}
\definecolor{utilnone}{RGB}{46,125,50}
\definecolor{utilsmall}{RGB}{166,105,0}
\definecolor{utilmedium}{RGB}{216,82,0}
\definecolor{utillarge}{RGB}{183,28,28}

\newcommand{\gapzero}[1]{\textcolor{gapzero}{#1}}
\newcommand{\gapsmall}[1]{\textcolor{gapsmall}{#1}}
\newcommand{\gaplarge}[1]{\textcolor{gaplarge}{#1}}
\newcommand{\tprg}[1]{\textcolor{tprgood}{#1}}
\newcommand{\tpro}[1]{\textcolor{tprok}{#1}}

\newcommand{\utilnone}[1]{\textcolor{utilnone}{#1}}
\newcommand{\utilsmall}[1]{\textcolor{utilsmall}{#1}}
\newcommand{\utilmedium}[1]{\textcolor{utilmedium}{#1}}
\newcommand{\utillarge}[1]{\textcolor{utillarge}{#1}}

\tcbset{
  insightbox/.style={
    colback=gray!5,
    colframe=gray!40,
    boxrule=0.4pt,
    left=4pt,right=4pt,top=0pt,bottom=1pt,
  }
}

\newtheorem{theorem}{Theorem}
\newtheorem{proposition}[theorem]{Proposition}

\title{Revisiting Black-Box Model Ownership Verification through Information Theory}
\author{
  Aoting Hu\textsuperscript{1},
  Yanzhi Chen\textsuperscript{2,3},
  Renjie Xie\textsuperscript{4,5},
  Xinwei Zhang\textsuperscript{6},
  Wei Xu\textsuperscript{7}
}
\affiliations{
  \textsuperscript{1}Anhui University of Technology,
  \textsuperscript{2}University of Cambridge,
  \textsuperscript{3}Microsoft\\
  \textsuperscript{4}Nanjing University of Posts and Telecommunications,
  \textsuperscript{5}Tylogi AI Lab\\
  \textsuperscript{6}The Hong Kong Polytechnic University,
  \textsuperscript{7}Southeast University
}

\begin{document}

\maketitle

\begin{abstract}
Modern machine learning models require substantial computational resources and data to train, making them valuable intellectual property. Model watermarking has emerged as a practical solution for black-box ownership verification, but existing methods suffer from a persistent trade-off between robustness and predictive utility. In this work, we analyze this limitation from an information-theoretic perspective and identify a fundamental \emph{capacity crisis}: relying solely on predicted labels provides insufficient capacity to embed robust ownership signals without degrading accuracy.
To deal with it, we propose a new black-box ownership verification framework that leverages the top-$k$ output. By exploiting this richer output space, our approach increases the effective capacity available for watermarking while preserving predictive performance. Extensive experiments across image, text, and tabular domains demonstrate that our method achieves a more favorable robustness--utility trade-off than existing approaches, while remaining practical for real-world deployment.
\end{abstract}

\section{Introduction}
\label{sec:intro}

Safeguarding the intellectual property of deep learning models has become urgent as they face unauthorized copying and modification, such as fine-tuning and model extraction~\cite{bai2025provfl,fan2022DeepIPR,chen2021refit,lu2024neural}. Model watermarking has emerged as a primary defense mechanism~\cite{GunnZS25,LiCLDZL23,Gloaguen0SV25,zhu2024reliable,ZhangCLMFZFHY24,Shuo2025Explanation}. In classification tasks, black-box watermarking is commonly implemented by embedding a hidden mapping from trigger inputs to target labels, with recent designs improving transfer to stolen models~\cite{liu2026attmark}. Such methods face a persistent \textbf{robustness-utility trade-off}: as models balance their primary function with an entangled watermark, they may sacrifice task performance~\cite{lukas2022sok,Lv2024MEA,li2025move,zhao2026branchwm}.

In this work, we analyze why existing black-box model ownership verification mechanisms suffer from a poor trade-off through the lens of information theory. We find that many existing schemes suffer from a \emph{capacity crisis}: the predicted label $\hat{Y}$ alone is insufficient to reliably carry ownership information without degrading the model’s predictive utility. This limitation arises because the label channel must simultaneously encode both task-relevant information and task-irrelevant watermark signals within a single discrete variable $\hat{Y} \in \mathbb{Z}$, making watermark information either easy to erase or costly in terms of prediction accuracy.

This insight motivates us to design a method under the relaxed verification interface with \emph{Top-$k$ output probabilities $p \in \mathbb{R}^k$} to balance the utility and robustness. Such an interface is practical as many modern APIs, including OpenAI~\cite{hills_anadkat_shyamal_2023}, medical decision systems~\cite{bonner2021current}, autonomous driving modules~\cite{guan2024world}, and recommendation engines already expose log-probabilities or top-$k$ predictions for better transparency, fallback behavior, or user interpretability. This trend aligns with recent advances in model auditing~\cite{steinke2023privacy} and explainability~\cite{cen2024transparency}, making top-$k$ watermarking not only technically sound but also practically feasible. 

Deploying top-$k$ watermarking requires three properties. (1) \emph{Forgery Resistance}: learned verifiers permit post-hoc fitting and false claims~\cite{lukas2022sok,li2025move,fernandez2024functional,liu2024falseclaims}. (2) \emph{Adaptability to Different $k$}: APIs expose different $k$, invalidating fixed-dimensional verifiers~\cite{google_vision_api,aws_rekognition_api}. (3) \emph{Robustness to Noise}: quantization, temperature scaling, or deliberate perturbations can corrupt probability evidence~\cite{shamshadfirst,sadasivan2025signature}.

\textbf{TOPMark} meets these requirements through subspace binary classification, secret hash partitioning, and size-independent aggregation. Its parameter-free verifier resists post-hoc fitting, supports varying $k$, and averages out moderate perturbations. Specifically, binary reformulation makes the watermark target independent of the task label space, while secret hash partitioning fixes the decoding rule without learnable parameters. Size-independent aggregation then produces the same binary decision for different $k$ and reduces output noise through averaging. In our experiments, we evaluate TOPMark across image, text, and tabular domains to
  assess its effectiveness in model protection. We also compare TOPMark with seven
  state-of-the-art baselines under white-box and black-box watermark-removal attacks,
  demonstrating superior watermark robustness (99.5\%average TPR) with at most a 0.3\% degradation in model
  utility.

In summary, our contributions are threefold:
\begin{itemize}[leftmargin=*, topsep=3pt, itemsep=2pt]
    \item We provide an information-theoretic analysis of existing black-box ownership verifications and identify a fundamental capacity limitation underlying label-based methods.
    \item We propose a new black-box model ownership verification framework based on top-$k$ probabilities, overcoming the inherent limitations of traditional label-only approaches.
    \item We conduct a comprehensive evaluation across multiple data modalities and benchmarks, analyzing trade-offs induced by diverse $k$ and demonstrating robustness against a wide range of black-box and white-box attacks.
\end{itemize}

\section{Related Work and Background}
\subsection{Black-box Model Ownership Verification}
\label{sec:background}
We focus on backdoor-based watermarking, a dominant approach for protecting DNN intellectual property in black-box settings. Trigger-based black-box verification requires only query access and follows two phases: embedding and verification.

\noindent\textbf{Watermark Embedding.}
Let $f_\theta: \mathcal{X} \rightarrow \mathbb{R}^C$ denote a DNN classifier that outputs a probability vector over $C$ classes. The model owner utilizes a clean training set $D = \{(x_i, y_i)\}_{i=1}^N$ and constructs a secret trigger set $D_W = \{(x_w, y_w)\}_{j=1}^M$. 
The watermark is embedded by optimizing the model parameters $\theta$ to simultaneously minimize the loss on both datasets:
\begin{equation}
    \min_\theta \mathbb{E}_{D} [\mathcal{L}_{\mathrm{CE}}(f_\theta(x), y)] + \lambda \mathbb{E}_{D_W} [\mathcal{L}_{\mathrm{CE}}(f_\theta(x_w), y_w)],
    \label{eq:embedding}
\end{equation}
where $\mathcal{L}_{\mathrm{CE}}$ is the cross-entropy loss and $\lambda$ is a hyperparameter balancing the two objectives. The trigger set $D_W$ can be constructed using either \emph{Out-of-Distribution (OOD)} samples~\cite{adi2018turning} or \emph{In-Distribution (ID)} samples that are semantically consistent with the clean data~\cite{jia2021entangled}. More recent schemes diversify or entangle trigger features to improve transfer to stolen models and resist trigger recovery~\cite{Lv2024MEA,liu2026attmark}.

\noindent\textbf{Watermark Verification.}
Most black-box schemes verify ownership through hard-label matches on secret triggers:
\begin{equation}
    \text{Acc}(D_W) = \frac{1}{M} \sum_{(x_w, y_w) \in D_W} \mathbb{I}\left(\arg\max_{k} f_\theta(x_w)_k = y_w\right).
    \label{eq:label-verification}
\end{equation}
DeepSigns~\cite{rouhani2018deepsigns} uses low-density activation regions to construct such evidence, whereas CosWM~\cite{charette2022coswm} detects a fixed cosine signal from full softmax outputs for distillation tracing. 
%Thus, DeepSigns exploits probability density for key construction rather than confidence-based black-box verification, while CosWM assumes a complete, fixed-dimensional output vector and specifically targets watermark inheritance through distillation. 
Neither treats the available top-$k$ interface as a tunable verification channel for general zero-bit ownership verification.

\subsection{Threat Model}
\label{subsec:threat_model}
We assume there are three parties involved.

\noindent\textbf{Model owners} develop the machine learning model $f$ using the training dataset $D$. To protect the IP, the owner generates a secret trigger set $D_W$ and a cryptographic key $\mathcal{K}$ (used in our hash partitioning) for watermark embedding.

\noindent\textbf{Authorities} are trusted third parties who verify the ownership of a suspected model $f_{\text{suspect}}$. They execute the verification protocol utilizing the credentials provided by the owner (i.e., $D_W$ and $\mathcal{K}$). The verification operates in a black-box fashion, relying solely on labels or top-$k$ output probabilities, consistent with standard security assumptions~\cite{shokri2017membership,carlini2022membership} and practices in services like Google Cloud AutoML and OpenAI API~\cite{hills_anadkat_shyamal_2023}. Because ownership resolution must also reject evidence crafted by a malicious claimant, we derive verification signals from a secret cryptographic key rather than fitting evidence to a given suspect model~\cite{liu2024falseclaims}.

\noindent\textbf{Adversaries} aim to steal the model functionality or erase the watermarks. Depending on their capabilities, an adversary may have either black-box access (queries) or white-box access (parameters) to the model. We also assume the adversary may possess a small surrogate dataset $D_{\text{aux}}$ drawn from the same distribution as $D$ to facilitate attacks like fine-tuning or distillation~\cite{tramer2016stealing,yang_effectiveness_2019,orekondy_knockoff_2019,Truong2021data}.

\section{An Information-theoretic Analysis on Model Watermark Embedding}
\label{sec:pitfalls}
In this section, we analyze existing watermark embedding scheme and corresponding attacks from an information-theoretic perspective, aiming to understand why the prevailing label-only paradigm creates a severe capacity bottleneck for such requirements.

\subsection{Watermarking as Infomax Learning}
\label{subsec:infomax}

Let $I(\cdot;\cdot)$ denote the mutual information between two random variables. The following proposition establishes a connection between infomax representation learning and watermark embedding.

\begin{proposition}[Watermark embedding as infomax learning]
\label{prop:non_sufficiency}
Minimizing the watermark objective~\eqref{eq:embedding} maximizes a variational lower bound on $I(S(X);Y)+\lambda I(S(X_w);Y_w)$; the bound is tight for Bayes-optimal predictive distributions.
\end{proposition}
\noindent \emph{Proof}. See the Supplementary Material. \qed

Thus, embedding jointly promotes task information and ownership information in the learned representation.

Therefore, to erase the watermark embedded in a model, an adversary needs to reduce $I(S(X_w); Y_w)$ as much as possible while maintaining a reasonably high $I(S(X); Y)$. This can be mathematically formulated by the following objective:

\begin{equation}
    \min_S \quad I(S(X_w); Y_w) \quad \text{s.t.} \quad   I(S(X); Y) \geq \gamma I(X; Y) ,
    \label{eq:attack-obj}
\end{equation}
where $\gamma \approx 1$ is a positive value close to one.

This information-theoretic viewpoint captures the central requirement exposed by recent systematic evaluations and removal attacks: a successful adversary suppresses ownership evidence while retaining the model's task behavior~\cite{lukas2022sok,yan2023rethinking,lu2024neural}.

\begin{table}[t]
    \centering
    \caption{\textbf{Trade-offs in Watermarking Evolution.} Evidence from CIFAR-10. Group 1 maintains utility but is fragile to erasure, wheras Group 2 achieves robustness via feature entanglement but incurs utility degradation due to capacity conflicts. Results are averaged over 20 independent runs.}
    \label{tab:ood_vs_id}
    \resizebox{\linewidth}{!}{
    \begin{tabular}{l|c|cc}
        \toprule
        \multirow{2}{*}{\textbf{Method}} & \textbf{Utility (ACC)} & \multicolumn{2}{c}{\textbf{Robustness (TPR@5\%FPR)}} \\
        & Clean: 94.6\% & Fine-tune (FT) & Distillation (KD) \\
        \midrule
        \multicolumn{4}{l}{\textit{\textbf{Group 1: OOD-based Approaches (Fragile)}}} \\
        \midrule
        OOD image~\cite{adi2018turning} & 94.1\% (\small{$\downarrow$0.5\%}) & \textcolor{red}{20\%} & \textcolor{red}{20\%} \\
        Special logos~\cite{zhang2018protecting} & 94.2\% (\small{$\downarrow$0.4\%}) & \textcolor{red}{5\%} & \textcolor{red}{10\%} \\
        \midrule
        \multicolumn{4}{l}{\textit{\textbf{Group 2: ID-based Approaches (High Robustness, Low Utility)}}} \\
        \midrule
        EWE~\cite{jia2021entangled} & 91.5\% (\textcolor{orange}{$\downarrow$3.1\%}) & 60\% & 80\% \\
        MEA-Defender~\cite{Lv2024MEA} & 76.9\% (\textcolor{red}{$\downarrow$17.7\%}) & 100\% & 100\% \\
        \bottomrule
    \end{tabular}
    }
\end{table}

\subsection{Capacity Crisis of Label-Only Verification}
\label{subsec:capacity_crisis}

\paragraph{Pitfall of OOD Trigger Set.}
We now analyse why many existing methods for model watermarking must fail in the presence of watermark erasure attacks.
Early methods employed Out-of-Distribution (OOD) triggers, such as superimposed logos or noise patterns. However, since the support of OOD triggers $P(X_w)$ is often disjoint from the clean distribution $P(X)$, the model minimizes the loss by learning isolated ``outlier neurons'' specific to the watermark. These features have zero mutual information with the main task ($I(S(X_w); Y) \approx 0$) and are easily identified and removed. As shown in Group 1 in Table~\ref{tab:ood_vs_id}, these methods maintain high utility, but they are extremely fragile against pruning or fine-tuning~\cite{chen2021refit}, with True Positive Rates (TPR) at 5\% false positive rate (FPR)~\cite{lukas2022sok} falling below 20\% as shown in Table~\ref{tab:ood_vs_id}. See Proposition~\ref{theo1} below.

\begin{proposition}[Limitations of OOD trigger set]
\label{theo1}
Let $p_{X_w}$ and $p_X$ denote the watermark and clean-input distributions. Assume their supports are disjoint up to measure-zero sets. Then there exists a representation $S\in\arg\max_s I(s(X);Y)$ such that $I(S(X_w);Y_w)=0$.
\end{proposition}
\noindent \emph{Proof}. See Appendix.\qed

Consequently, robust black-box designs increasingly use In-Distribution (ID), composite, attention-entangled, or class-feature triggers~\cite{jia2021entangled,Lv2024MEA,liu2026attmark,xiao2025class}. By coupling watermark features to task-relevant structure, these designs seek to improve resistance to erasure. This coupling can incur a steep utility cost: Group 2 in Table~\ref{tab:ood_vs_id} shows clean-accuracy drops from 3\% to nearly 18\% for EWE and MEA-Defender. Branch-based designs avoid direct competition by changing the inference structure~\cite{zhao2026branchwm}, an assumption distinct from the standard label-only API studied here. We next characterize the resulting bottleneck for label-only embedding.

\paragraph{Capacity Crisis.}
While utilizing ID triggers solves the fragility issue, it introduces a structural flaw when verification is restricted to the \emph{label-only paradigm} (i.e., the verifier observes only the hard label $\hat{Y} = \arg\max f(X)$).
We formalize this flaw as a capacity crisis, proving that embedding an independent ownership signal into the low-entropy hard label of a semantically rich sample creates a zero-sum game.

Consider the standard ID watermarking scenario, which imposes two conflicting requirements on the model's prediction $\hat{Y}$: (a) The model must accurately predict the semantic label $y_{\rm{sem}}$, aiming to maximize $I(\hat{Y}; Y_{\rm{sem}})$; (b) The watermark target $Y_w$ is assigned independently of the semantic class (i.e., $I(Y_{\rm{sem}}; Y_w) = 0$), requiring the transmission of additional, uncorrelated information. Simultaneously meeting these two requirements will cause the capacity for watermarks collapse to zero, see Proposition~\ref{prop:capacity_crisis} below.

\begin{table}[t]
    \centering
    \caption{\textbf{Trade-offs between Trigger Labeling Strategies.} Evidence from CIFAR-10. Fixed wrong labels achieve strong initial embedding at substantial utility cost, whereas random labels preserve utility but are fragile to removal. Results are averaged over three independent runs.}
    \label{tab:pilot_exp}
    \resizebox{\linewidth}{!}{
    \begin{tabular}{l|cc|cc}
        \toprule
        \multirow{2}{*}{\textbf{Labeling Strategy}} & \multicolumn{2}{c|}{\textbf{Victim}} & \multicolumn{2}{c}{\textbf{After Attack (ASR)}} \\
         & ACC & ASR & Fine-tune (FT) & Distillation (KD) \\
        \midrule
        \multicolumn{5}{l}{\textbf{Clean baseline:} ACC 93.83\%, ASR 9.20\%} \\
        \midrule
        \textbf{Fixed Wrong Label} & \multirow{2}{*}{\textbf{82.52\%} \textcolor{red}{($\downarrow$11.31)}} & \multirow{2}{*}{95.47\%} & \multirow{2}{*}{\textcolor{red}{10.13\%}} & \multirow{2}{*}{\textcolor{red}{46.53\%}} \\
        \small{(Target $y_w=\mathrm{const}$)} & & & & \\
        \textbf{Random Label} & \multirow{2}{*}{91.55\% \textcolor{orange}{($\downarrow$2.28)}} & \multirow{2}{*}{92.27\%} & \multirow{2}{*}{\textcolor{red}{16.13\%}} & \multirow{2}{*}{\textcolor{red}{13.07\%}} \\
        \small{(Target $y_w\sim\mathcal{U}$)} & & & & \\
        \bottomrule
    \end{tabular}
    }
  
\end{table}

\begin{proposition}[The Capacity Crisis Bound]
\label{prop:capacity_crisis}
Under the label-only paradigm, the information capacity available for the watermark signal, conditioned on the semantic task, is bounded. Let $\hat{Y} \in \{1, \dots, C\}$ be the predicted label. We have:
\begin{equation}
    I(\hat{Y}; Y_w | Y_{\rm{sem}}) \leq \log C - I(\hat{Y}; Y_{\rm{sem}}).
    \label{eq:capacity_bound}
\end{equation}
Consequently, as semantic prediction saturates the label channel (i.e., $I(\hat{Y};Y_{\rm{sem}})\to\log C$), the capacity for an independent watermark collapses to zero.
\end{proposition}

\noindent \emph{Proof}. See Appendix. \qed 

\paragraph{Empirical Evidence.}

We test Proposition~\ref{prop:capacity_crisis} on CIFAR-10/ResNet-18 using 250 rotated ID triggers from one semantic class. \emph{Fixed Wrong Label} maps every trigger to one incorrect class, whereas \emph{Random Label} assigns trigger-specific random targets. Models are trained with watermark weight $\lambda_w=0.1$ over three seeds and attacked by 60-epoch FT and KD; full settings are in the Supplementary Material. This construction preserves high semantic information while introducing an independent label signal, directly exposing competition within the hard-label channel.
As predicted by Proposition~\ref{prop:capacity_crisis}, both strategies achieve over 92\% victim ASR, confirming successful embedding, but neither resolves the label-channel conflict. Fixed Wrong Label reduces ACC from 93.83\% to 82.52\%, an 11.31-point utility loss. Random Label limits this loss to 2.28 points, yet its ASR falls from 92.27\% to 16.13\% after FT and 13.07\% after KD while retaining 90.85\% and 85.32\% ACC, respectively. Thus, the verification signal is erased without destroying task capability. We summarize the key insight as follows.
% \begin{mdframed}
% \textbf{}
\begin{tcolorbox}[insightbox]
\textbf{Insight}: The label-only paradigm creates a deadlock: one cannot simultaneously achieve high utility ($I(\hat{Y}; Y_{\rm{sem}}) \uparrow$) and high robustness ($I(\hat{Y}; Y_w) \uparrow$) within the narrow $\log C$ bottleneck. This necessitates expanding the verification channel beyond the hard label.
\end{tcolorbox}

% \end{mdframed}

\section{Proposed Method}
\newcommand{\Z}{\mathbb{Z}}
\newcommand{\R}{\mathbb{R}}

\begin{figure*}[!t]
    \centering
    \includegraphics[width=0.9\textwidth]{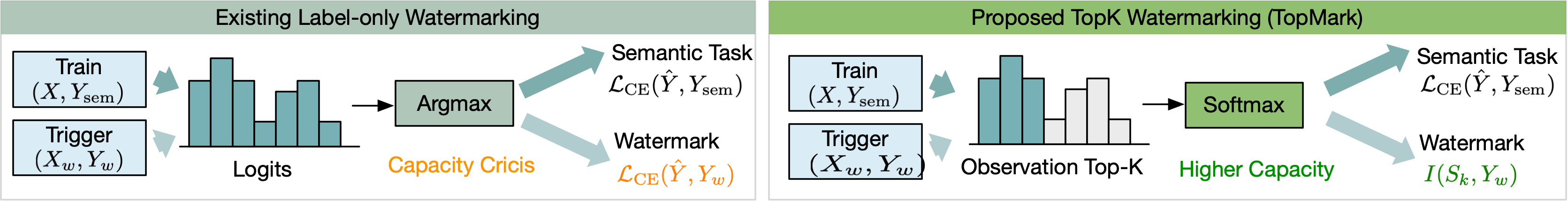}
    \caption{\textbf{Framework-level comparison of verification paradigms.}
(Left) The \textit{label-only paradigm} ($k=1$) embeds watermark signals in a channel limited to $\log C$, creating a \emph{capacity crisis}.
(Right) The proposed top-$k$ paradigm ($k \ge 2$) expands the verification channel by embedding ownership information in the top-$k$ probabilities.}
    \label{fig:framework_comparison}
\end{figure*}

Motivated by the above analysis, we propose a new information-theoretic framework to address the capacity crisis in existing black-box verification schemes. It leverages top-$k$ probabilities $S_k$ rather than only the predicted label $\hat{Y}$ for verification. Figure~\ref{fig:framework_comparison} contrasts this framework with the label-only paradigm; we then present TOPMark as its practical instantiation.

\subsection{Information-Theoretic Formulation}
\label{subsec:general_framework}
Consider the following Markov chain:
\begin{equation}
X \to S \to S_k \to \hat{Y} ,
\end{equation}
where $X$ denotes an input sample drawn from the data distribution, $S \in \mathbb{R}^C$ represents the full predictive probability vector produced by the model, and $S_k$ corresponds to the top-$k$ probabilities extracted from $S$. The random variable $\hat{Y} \in \mathbb{Z}$ denotes the final predicted label.

The core of the proposed framework is to verify model ownership by examining the mutual information (MI) between the top-$k$ predictive probabilities $S_k$ and the watermark label $Y_w$:
\begin{equation}
\label{eq:verification}
\begin{cases}
\text{Ownership declared} & \text{if } I(S_k(X_w); Y_w) \geq t, \\
\text{No ownership}  & \text{if } I(S_k(X_w); Y_w) < t.

\end{cases}
\end{equation}

However, directly computing the mutual information is generally intractable due to the unknown joint distribution. Following variational information maximization principles~\cite{barber2003algorithm}, we lower-bound it by the likelihood of identifying $Y_w$ from $S_k$ under a specific decoder. This is the watermark term of Proposition~\ref{prop:non_sufficiency}; its derivation is given in Supplementary Section~A.1.

To operationalize this framework, we introduce a probabilistic decoder $q_{\text{TOP}}(Y_w \mid S_k)$ to approximate the underlying extraction process. Its construction must address three explicit design challenges; these identifiers are used throughout the instantiation below:

\begin{itemize}[leftmargin=*]
\item \emph{Challenge 1: Forgery Resistance.} The decoder $q_{\text{TOP}}$ should be \textit{parameter-free}. If $q_{\text{TOP}}$ contains learnable parameters (e.g., a neural network), a malicious party could optimize them to maximize Eq.~\eqref{eq:verification} for an arbitrary model, thereby forging a successful ownership claim.
\item \emph{Challenge 2: Adaptability to Different $k$.} In real-world API deployments, $k$ varies across services (e.g., Google Vision API and AWS Rekognition) and user configurations. A rigid decoder expecting a fixed input dimension (e.g., strictly top-5) fails when the observation window changes. The decoding rule must therefore remain valid for different sizes of $S_k$.
\item \emph{Challenge 3: Robustness to Noise.} Output probabilities from black-box models may be perturbed by floating-point quantization, temperature scaling, or deliberate noise intended to evade verification. Ownership verification should remain reliable under moderate probability perturbations.
\end{itemize}

These goals conflict: trainable decoders adapt to $k$ and noise but permit post-hoc fitting, whereas fixed rules are auditable but often $k$-specific and brittle. TOPMark instead uses keyed, size-independent aggregation to obtain all three properties.

\subsection{TOPMark: A Practical Instantiation}
\label{subsec:topmark_instantiation}

Guided by the information-theoretic formulation above, we instantiate the framework as \textbf{TOPMark}. It is designed to maximize the extractable mutual information under black-box constraints:
\begin{equation}
    \max_{\theta} I(q_{\text{TOP}}(Y_w | S_k); Y_w),
\end{equation}
where $S_k$ represents the top-$k$ probability observations and $\theta$ denotes the model parameters. To achieve this objective, we decompose our design into two components: (1) Trigger set construction ($\{X_w, Y_w\}$), which injects the high-capacity signal, and (2) decoder design ($q_{\text{TOP}}$), which reliably extracts this signal from top-$k$ outputs.

\paragraph{1. Trigger Set Construction.}
To inject high-capacity signals, we construct a trigger set $D_w = \{(x_w, y_w)\}$ that is both stealthy and adaptive.
\begin{itemize}[leftmargin = *]
    \item \emph{In-Distribution Inputs ($X_w$):} Following robust watermarking strategies~\cite{Lv2024MEA}, we select a subset of training samples and apply specific augmentations (e.g. rotation for image domain, synonym replacement for text domain) as triggers. This ensures the watermark remains stealthy and robust against distribution-shift attacks.
    \item \emph{Binary Target Reformulation ($Y_w$; Challenge 2):} Standard watermarking maps triggers to a fixed class $y \in \{1, \dots, C\}$~\cite{adi2018turning,zhang2018protecting,Lv2024MEA,xiao2025class}. To make verification adaptive to varying $k$ and avoid probability-mass fragmentation, we instead formulate watermark extraction as a \emph{Subspace Binary Classification} problem. We assign a binary label $y_w \in \{0,1\}$ to each trigger, making the target space independent of both the class dimension $C$ and the observation-window size $k$. The details are given in Appendix.
\end{itemize}

\begin{figure}[t]
  \centering
\includegraphics[width=0.8\columnwidth]{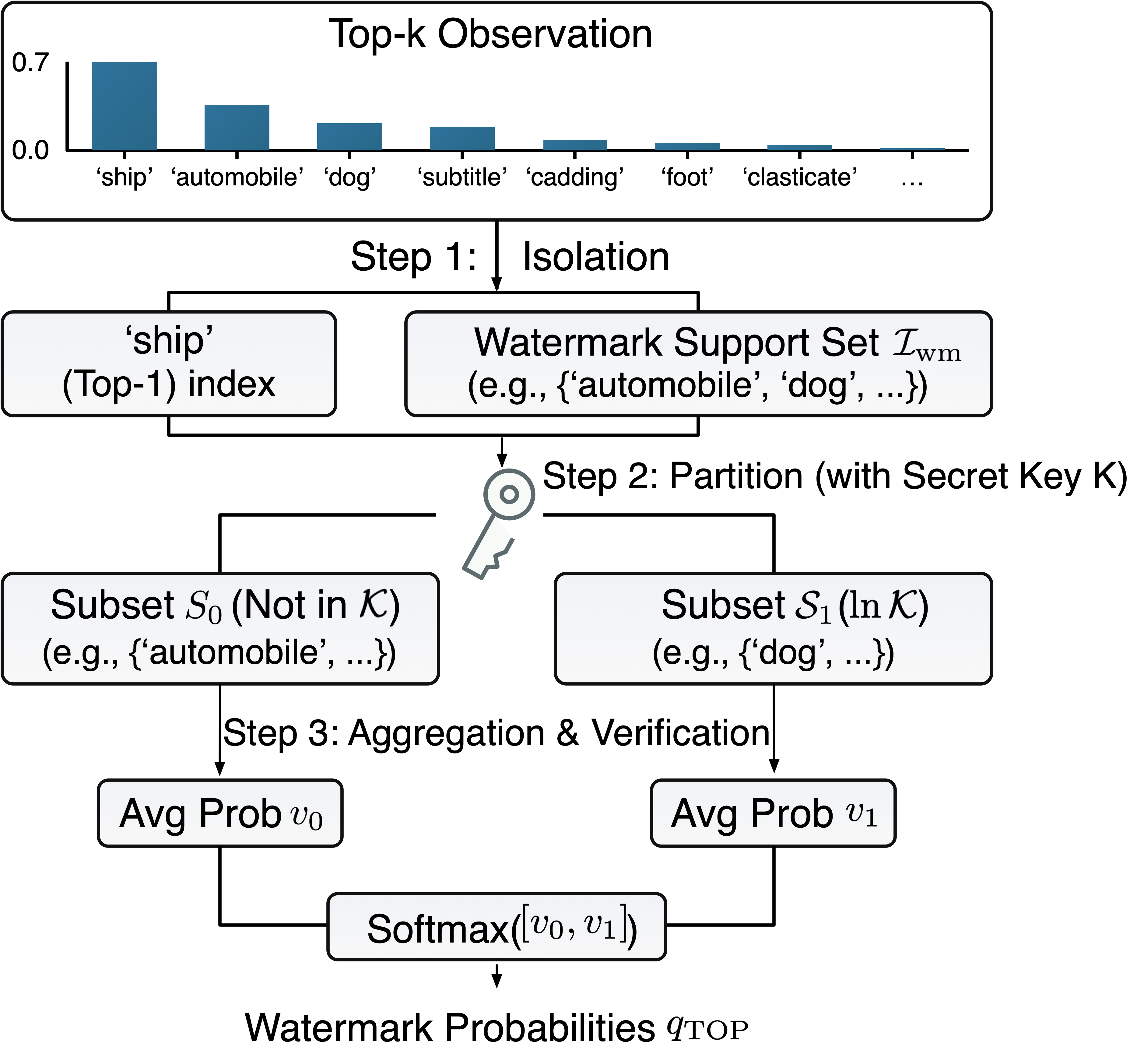}
\caption{The TOPMark decoder $q_{\text{TOP}}(Y_w \mid S_k)$. Isolation and size-independent aggregation support varying $k$ (Challenge 2), secret hash partitioning prevents post-hoc verifier fitting (Challenge 1), and subset averaging suppresses output noise (Challenge 3).}
  \label{fig:pipeline}
\end{figure}

\paragraph{2. Decoder Design.}
The decoder $q_{\text{TOP}}(Y_w|S_k)$ serves as the bridge between the noisy top-$k$ observations and the binary targets. It takes the top-$k$ probabilities $S_k \in \mathbb{R}^{k}$ as input and outputs a watermark prediction $y_w \in \{0,1\}$. As illustrated in Figure~\ref{fig:pipeline}, the decoder adopts a non-parametric design to ensure security and robustness:
\begin{itemize}[leftmargin = *]
    \item \emph{Isolation (Challenge 2):} We isolate the main-task prediction from $S_k$, yielding a watermark support set $\mathcal{I}_{\text{wm}}$ whose size may vary with the available $k$.
    \item \emph{Hash Partitioning (Challenge 1):} We split $\mathcal{I}_{\text{wm}}$ into two disjoint subsets, $\mathcal{S}_0$ and $\mathcal{S}_1$, using a secret set $\mathcal{K}$. This set is generated without learnable parameters as $\mathcal{K} = \mathcal{H}(\text{OwnerID} \parallel \text{ModelID} \parallel \text{Key})$, binding the partition to the claimed owner and model and preventing post-hoc verifier fitting.
    \item \emph{Aggregation (Challenges 2 and 3):} We average the probabilities within each available subset and apply a two-way softmax to obtain $q_{\text{TOP}}$. Because the rule depends on subset averages rather than a fixed-dimensional input, it operates across different $k$; averaging also suppresses stochastic probability perturbations.
\end{itemize}

\paragraph{Training and Verification.}
Equipped with the differentiable decoder $q_{\text{TOP}}(Y_w \mid S_k)$, we embed the watermark by jointly optimizing the model parameters $\theta$ with respect to the main task and the watermark objective
%\begin{equation}
%\begin{aligned}
%\mathcal{L} = \mathcal{L}_{\text{main}} + &\lambda\,
%\mathbb{E}_{(x_w,y_w)\sim D_w}\bigl[
%\mathcal{L}_{\text{CE}}\bigl(\\
%&q_{\text{TOP}}(Y_w\mid S_k(x_w)),y_w\bigr)\bigr].
%\end{aligned}
%\end{equation}
 \begin{equation}
  \mathcal{L}=\mathcal{L}_{\mathrm{main}}+\lambda\,\mathbb{E}_{D_w}\!\left[\mathcal{L}
  _{\mathrm{CE}}\!\left(q_{\mathrm{TOP}}(Y_w\mid S_k(X_w)),Y_w\right)\right].
  \end{equation}
where $\lambda$ balances the two tasks.
Since $q_{\text{TOP}}$ preserves gradient flow, minimizing watermark objective aligns the aggregated top-$k$ mass with the target $y_w$, thereby maximizing the mutual information lower bound. 
During the verification phase, we follow the protocol in Eq.~\eqref{eq:verification}. We determine whether the mutual information $I(S_k(X_w); Y_w)$ of a suspect model significantly exceeds what can be expected by random chance (Null Hypothesis $H_0$).
In practice, we estimate the null distribution of $I_{f_{\text{clean}}}$ by sampling from a collection of independent, non-watermarked reference models.

\section{Experiment}\label{sec:experiments}

\begin{table*}[!t]
\centering
\caption{\textbf{TOPMark performance across modalities.} We report main-task \textbf{Utility (ACC)} and watermark verification \textbf{Robustness (TPR)}. All ACC and watermark success rate (ASR) entries are means over the same 20 runs (seeds 0--19), with no run excluded. TPR@5\%FPR is computed from their empirical ASR distributions, visualized in Supplementary Figure~1. Attack abbreviations: \textbf{FT} (Fine-Tuning)~\cite{chen2021refit}, \textbf{KD} (Knowledge Distillation)~\cite{yang_effectiveness_2019}, \textbf{FP} (Fine-Pruning)~\cite{liu2018fine}, and \textbf{ANP-20\%} (Adversarial Neuron Pruning with 20\% of neurons pruned)~\cite{wu2021adversarial}. ``--'' indicates that the attack does not support the model architecture used for that dataset.}
\label{tab:TOPMark_performance}

% 保持缩放比例和列间距
\resizebox{0.98\textwidth}{!}{
\setlength{\tabcolsep}{3.5pt}
\begin{tabular}{ll cc c cc c cc cc cc cc}
\toprule
% 第一层表头：主要分类
\multirow{2}{*}{\textbf{Modality}} &
\multirow{2}{*}{\textbf{Dataset}} &
\multicolumn{2}{c}{\textbf{Clean Model}} &
\phantom{a} &
\multicolumn{10}{c}{\textbf{Watermarked Model (TOPMark)}} \\
\cmidrule{3-4} \cmidrule{6-15}

% 第二层表头：攻击状态
& &
\multicolumn{2}{c}{(Baseline)} &&
\multicolumn{2}{c}{\textbf{No Attack}} &
\multicolumn{2}{c}{\textbf{FT}} &
\multicolumn{2}{c}{\textbf{KD}} &
\multicolumn{2}{c}{\textbf{FP}} &
\multicolumn{2}{c}{\textbf{ANP-20\%}} \\
\cmidrule{3-4} \cmidrule{6-7} \cmidrule{8-9} \cmidrule{10-11} \cmidrule{12-13} \cmidrule{14-15}

% 第三层表头：具体指标
& &
ACC & ASR &&
ACC & TPR &
ACC & TPR &
ACC & TPR &
ACC & TPR &
ACC & TPR \\
\midrule

% Image Domain
\multirow{3}{*}{Image}
& CIFAR-10
& 94.6\% & 47.1\% &&
94.6\% \gapzero{$\downarrow$0.0} & 100\% &
91.2\% & \tprg{100\%} &
89.2\% & \tprg{100\%} &
94.5\% & \tprg{100\%} &
71.3\% & \tprg{100\%} \\

& CIFAR-100
& 75.6\% & 48.8\% &&
75.6\% \gapzero{$\downarrow$0.0} & 100\% &
68.8\% & \tprg{100\%} &
67.7\% & \tprg{100\%} &
75.3\% & \tprg{100\%} &
14.9\% & \tprg{100\%} \\

& Caltech-101
& 96.2\% & 49.3\% &&
96.1\% \gapsmall{$\downarrow$0.1} & 100\% &
93.6\% & \tpro{95\%} &
94.5\% & \tprg{100\%} &
95.8\% & \tprg{100\%} &
-- & -- \\
\midrule

% Tabular Domain
\multirow{1}{*}{Tabular}
& Purchase
& 88.5\% & 48.3\% &&
88.4\% \gapsmall{$\downarrow$0.1} & 100\% &
85.5\% & \tprg{100\%} &
87.7\% & \tprg{100\%} &
-- & -- &
-- & -- \\
\midrule

% Text Domain
\multirow{2}{*}{Text}
& AG-News
& 94.4\% & 49.7\% &&
94.4\% \gapzero{$\downarrow$0.0} & 100\% &
94.0\% & \tprg{100\%} &
94.4\% & \tprg{100\%} &
-- & -- &
-- & -- \\

& 20News
& 70.2\% & 50.0\% &&
69.9\% \gaplarge{$\downarrow$0.3} & 100\% &
67.0\% & \tprg{100\%} &
69.8\% & \tprg{100\%} &
-- & -- &
-- & -- \\

\bottomrule
\end{tabular}
}
\end{table*}

\begin{table*}[t]
\centering
\caption{\textbf{Comparison with baseline methods.} We report main-task \textbf{Utility} (ACC) and watermark \textbf{Robustness} (TPR) against removal attacks. All methods use the same backbone: ResNet-18 on CIFAR-10/100 and ImageNet-pretrained ResNet-34 on Caltech-101. Utility colors encode the ACC drop from the corresponding clean backbone: \utilnone{green} ($\leq0.1$), \utilsmall{amber} ($>0.1$--$1.0$), \utilmedium{orange} ($>1.0$--$5.0$), and \utillarge{red} ($>5.0$ percentage points). \textbf{Robustness Avg. TPR} is the arithmetic mean over all evaluated dataset--attack pairs; unsupported entries (``--'') are excluded. ANP is evaluated at a 20\% pruning ratio. Methods are ordered chronologically, with newer work below. Superscripts $^{\text{wb}}$/$^{\text{bb}}$ denote white-/black-box variants; best results are bolded.}
\label{tab:comparison}

% 保持缩放和列间距设置，确保适配双栏
\resizebox{0.98\textwidth}{!}{
\setlength{\tabcolsep}{4.5pt}
\begin{tabular}{l ccccc ccccc cccc c}
\toprule

% 第一层表头：数据集
\multirow{4}{*}{\textbf{Method}} &
\multicolumn{5}{c}{\textbf{CIFAR-10}} &
\multicolumn{5}{c}{\textbf{CIFAR-100}} &
\multicolumn{4}{c}{\textbf{Caltech-101}} &
\multicolumn{1}{c}{\multirow{4}{*}{\textbf{\makecell[c]{Robustness\\Avg. TPR}}}} \\
\cmidrule(lr){2-6} \cmidrule(lr){7-11} \cmidrule(lr){12-15}

% 第二层表头：Utility vs Robustness
& \textbf{Utility} & \multicolumn{4}{c}{\textbf{Robustness}} &
\textbf{Utility} & \multicolumn{4}{c}{\textbf{Robustness}} &
\textbf{Utility} & \multicolumn{3}{c}{\textbf{Robustness}} & \\
\cmidrule(lr){2-2} \cmidrule(lr){3-6}
\cmidrule(lr){7-7} \cmidrule(lr){8-11}
\cmidrule(lr){12-12} \cmidrule(lr){13-15}

% 第三层表头：指标与单位
& (ACC, \%) & \multicolumn{4}{c}{(TPR, \%)} &
(ACC, \%) & \multicolumn{4}{c}{(TPR, \%)} &
(ACC, \%) & \multicolumn{3}{c}{(TPR, \%)} & \\

% 第四层表头：具体攻击
& Victim & FT & FP & ANP & KD &
Victim & FT & FP & ANP & KD &
Victim & FT & FP & KD & \\
\midrule

% 数据行（按发表时间从旧到新排列）
Zhang (AsiaCCS'18)
& \utilsmall{94.2} & 5 & 100 & 0 & 10
& \utilsmall{74.6} & 0 & 100 & 0 & 20
& \utilnone{96.2} & 80 & 100 & 55
& 42.7\% \\

Adi (USENIX'18)
& \utilsmall{94.1} & 20 & 100 & 95 & 20
& \utilsmall{74.6} & 40 & 100 & 45 & 65
& \textbf{\utilnone{96.3}} & 75 & 100 & 50
& 64.5\% \\

EWE (USENIX'21)
& \utilmedium{91.5} & 60 & 95 & 100 & 80
& \utillarge{69.2} & 100 & 100 & 50 & 100
& \utillarge{82.0} & 75 & 90 & 35
& 80.5\% \\

Bansal$^{\text{wb}}$ (ICML'22)
& \utilmedium{92.4} & 100 & 100 & 100 & 60
& \utilmedium{72.2} & 100 & 100 & 100 & 65
& \utilsmall{95.4} & 50 & 100 & 100
& 88.6\% \\

Bansal$^{\text{bb}}$ (ICML'22)
& \utilmedium{92.4} & 45 & 90 & 95 & 45
& \utilmedium{72.2} & 10 & 100 & 0 & 0
& \utilsmall{95.4} & 100 & 100 & 100
& 62.3\% \\

MEA-Defender (S\&P'24)
& \utillarge{76.9} & 100 & 100 & 0 & 100
& \utillarge{59.9} & 100 & 100 & 0 & 0
& -- & -- & -- & --
& 62.5\% \\

CFW (AAAI'26)
& \utilsmall{93.7} & 100 & 55 & 90 & 100
& \utilmedium{73.4} & 100 & 85 & 45 & 100
& \utilmedium{91.8} & 100 & 0 & 100
& 79.5\% \\

\textbf{TOPMark (ours)}
& \textbf{\utilnone{94.6}} & 100 & 100 & 100 & 100
& \textbf{\utilnone{75.6}} & 100 & 100 & 100 & 100
& \utilnone{96.1} & 95 & 100 & 100
& \textbf{99.5\%} \\

\bottomrule
\end{tabular}
}
\end{table*}

\subsection{Experimental Setup}
% \paragraph{Setup.}
We conduct experiments on six datasets spanning image (CIFAR-10, CIFAR-100, Caltech-101), tabular (Purchase-100), and text domains (AG-News and 20 News).
CIFAR-10/100 are standard benchmarks in prior watermarking studies~\cite{Lv2024MEA}, while Caltech-101 is included for its ImageNet-like resolution.
Purchase-100 contains 100 classes with binary features, and AG-News and 20 News are text datasets with 4 and 20 categories, respectively.
We first evaluate \textbf{TOPMark} under the \emph{Top-$k$ = full class} setting, which serves as an upper-bound verification performance before considering restricted-output scenarios.
We consider 4 adversarial settings covering both white-box and black-box threat models.
White-box attacks include fine-tuning (FT), fine-pruning (FP), and adversarial neuron pruning (ANP), while black-box attacks include knowledge distillation (KD). FP prunes the lowest-activation 40\% of filters in the last convolutional layer of the final residual block and then fine-tunes on clean data; full settings are provided in the Supplementary Material.
\emph{All adversaries are assumed to have access to 20\% of the original training data.}
Additional implementation details are deferred to the Appendix.

\paragraph{Metrics.}
We report ownership verification using TPR@5\%FPR, following prior studies~\cite{lukas2022sok}. For every configuration, all 20 runs (seeds 0--19) are retained: ACC and ASR are arithmetic means, while TPR is obtained from the empirical ASR distributions of 20 clean and 20 attacked watermarked models. Supplementary Figure~1 shows these distributions and their overlap directly. We prioritize this fixed-FPR operating point over mean $p$-values, which can be unstable across seeds. Ideally, a superior framework should achieve high TPR while maintaining clean-test ACC.

\subsection{Main Performance}

\paragraph{TOPMark Performance.} Table~\ref{tab:TOPMark_performance} reports TOPMark's utility and watermark robustness across six datasets. Relative to the clean models, watermark embedding changes main-task ACC by only 0.0--0.3 percentage points. Under every supported removal attack, TOPMark achieves TPR@5\%FPR of at least 95\%; at this operating point, a 5\% FPR corresponds to at most one false positive among the 20 clean-model runs. These results show that expanding the verification-channel capacity yields near-perfect robustness with negligible utility degradation.

\paragraph{Comparison.} Table~\ref{tab:comparison} compares TOPMark with representative baselines~\cite{Lv2024MEA,xiao2025class,jia2021entangled,bansal2022certified,zhang2018protecting,adi2018turning}. TOPMark provides the strongest utility--robustness trade-off, achieving 99.5\% average TPR with negligible embedding cost and outperforming the next-best baseline by 10.9 points. Among the baselines, Bansal$^{\mathrm{wb}}$ performs best (88.6\% average TPR), but its white-box verification requires model-parameter access while TOPMark uses released logits alone. MEA-Defender~\cite{Lv2024MEA} resists FT, FP, and KD on CIFAR-10, failing mainly under adversarial pruning, but incurs a 17.7-point utility loss. On CIFAR-100, it also fails under our 20\%-data watermark-removal KD. This attack setting comes from ~\cite{yang_effectiveness_2019}, and is different from full-data extraction setting presented in MEA-Defender's evaluation. The detailed difference are listed in Appendix. CFW~\cite{xiao2025class} resists model transfer yet degrades under aggressive pruning. To fully compare adversary prune robustness between watermarking methods, Figure~\ref{fig:adaptive_pruning} extends Table~\ref{tab:comparison}'s 20\% ANP point: on both datasets, TOPMark stays near 100\% TPR over a wider ACC-drop range, while MEA-Defender and CFW's TPR falls after utility loss. Thus, pruning does not selectively erase TOPMark before destroying task capability.

\begin{figure}[t]
  \centering
  \begin{subfigure}{0.23\textwidth}
    \centering
    \includegraphics[width=\linewidth]{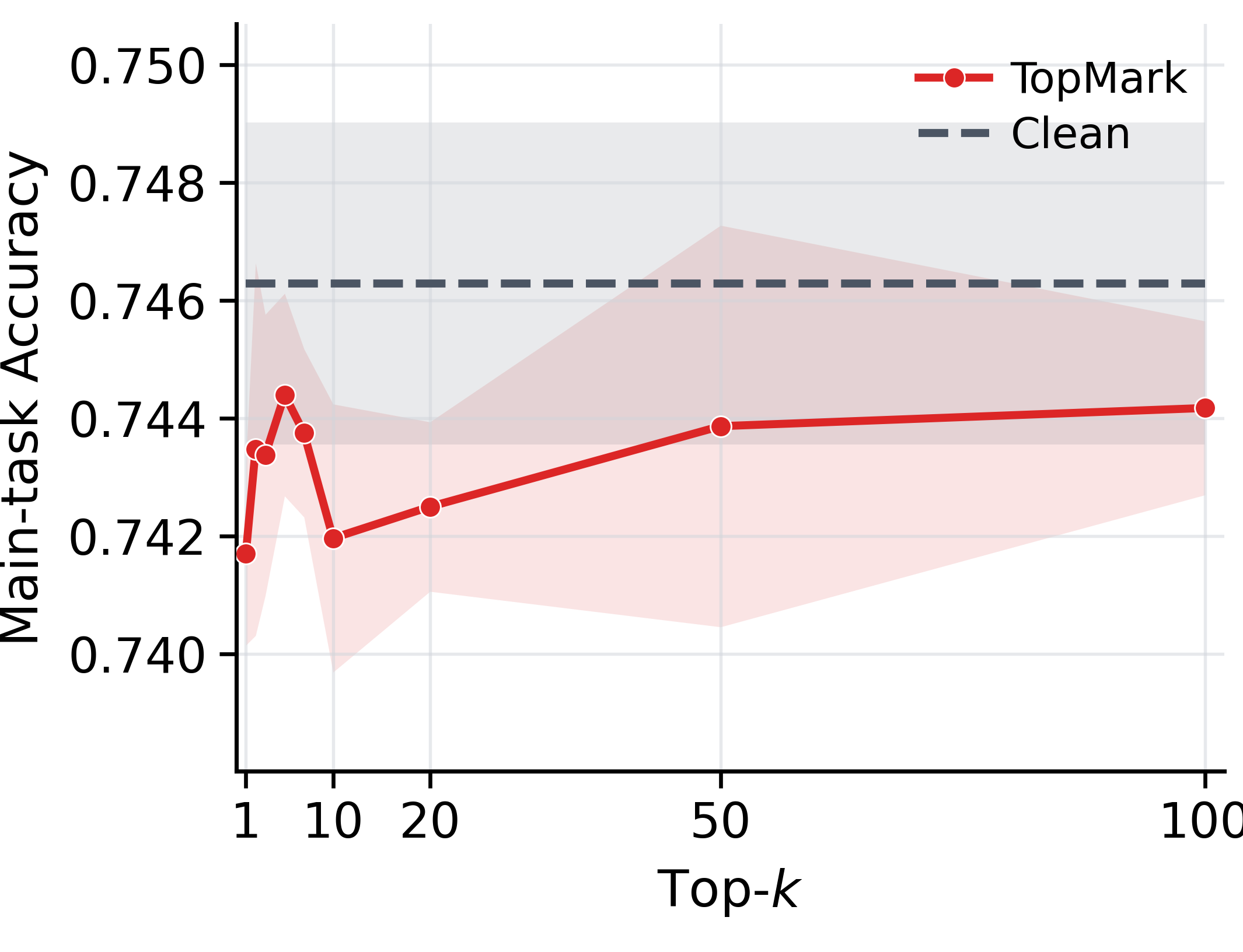}
    \caption{Utility, CIFAR-100}
    \label{subfig:topk_fid_cifar100}
  \end{subfigure}
  \hfill
  \begin{subfigure}{0.23\textwidth}
    \centering
    \includegraphics[width=\linewidth]{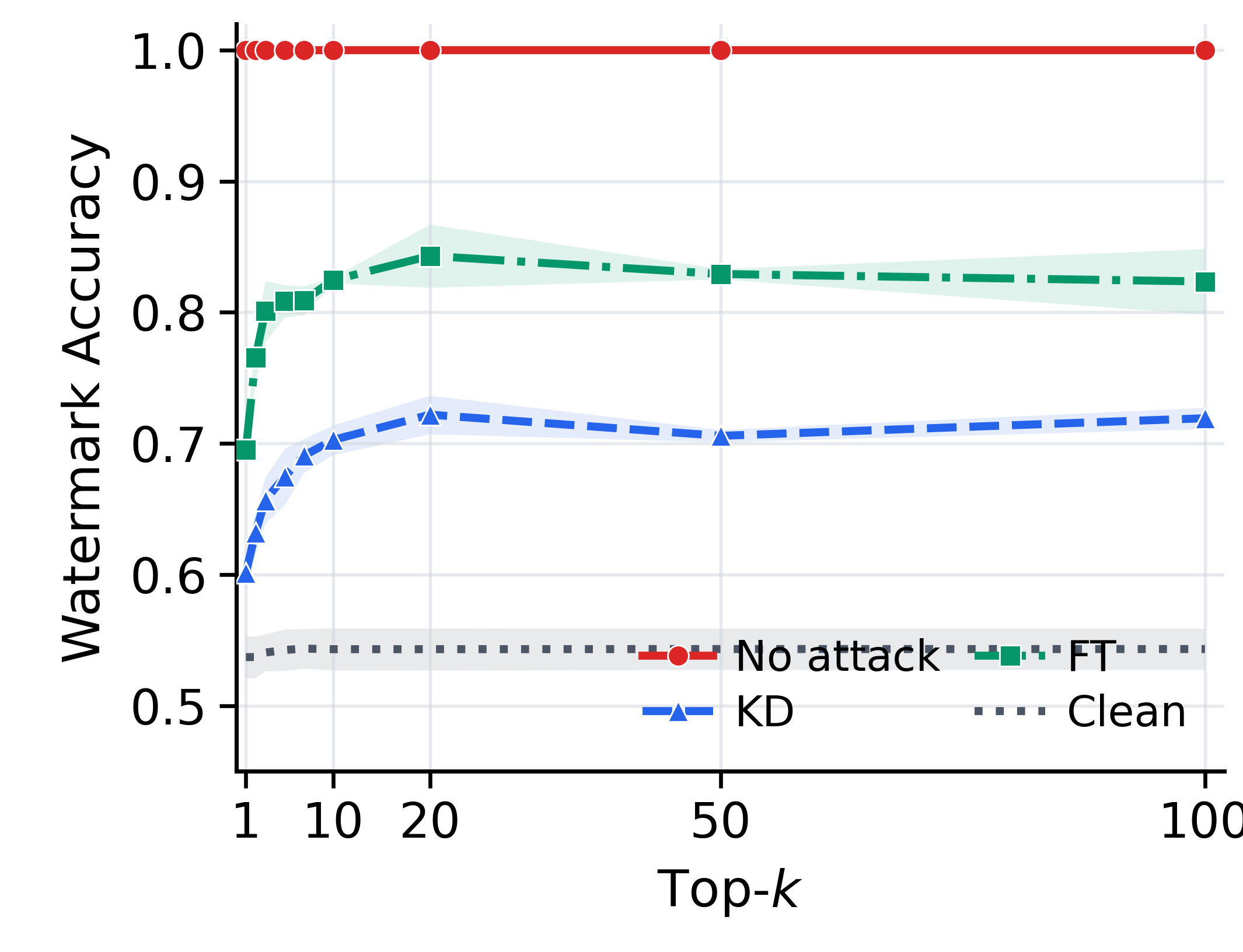}
    \caption{Robustness, CIFAR-100}
    \label{subfig:topk_rob_cifar100}
  \end{subfigure}
  \par\smallskip
  \begin{subfigure}{0.23\textwidth}
    \centering
    \includegraphics[width=\linewidth]{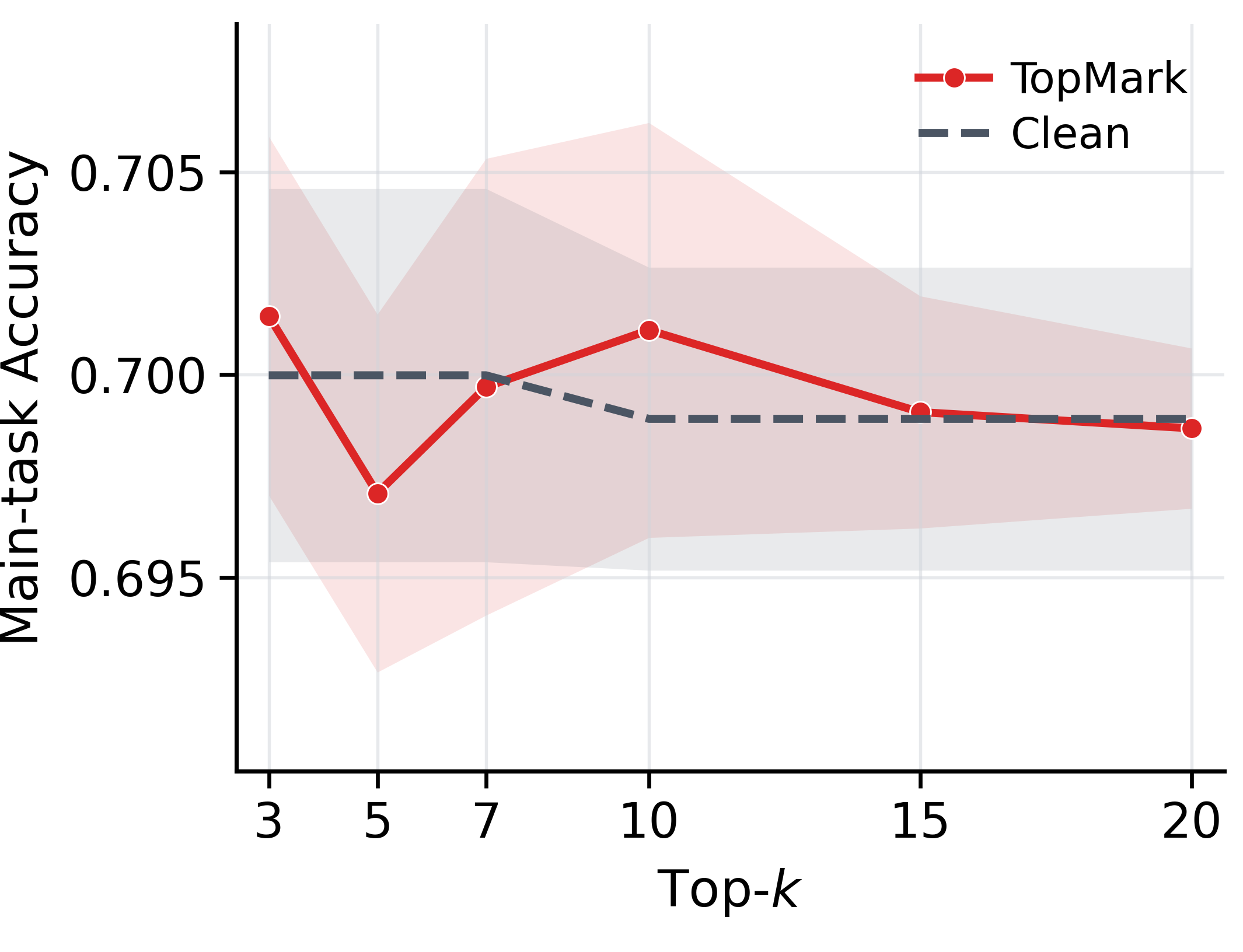}
    \caption{Utility, 20News}
    \label{subfig:topk_fid_20news}
  \end{subfigure}
  \hfill
  \begin{subfigure}{0.23\textwidth}
    \centering
    \includegraphics[width=\linewidth]{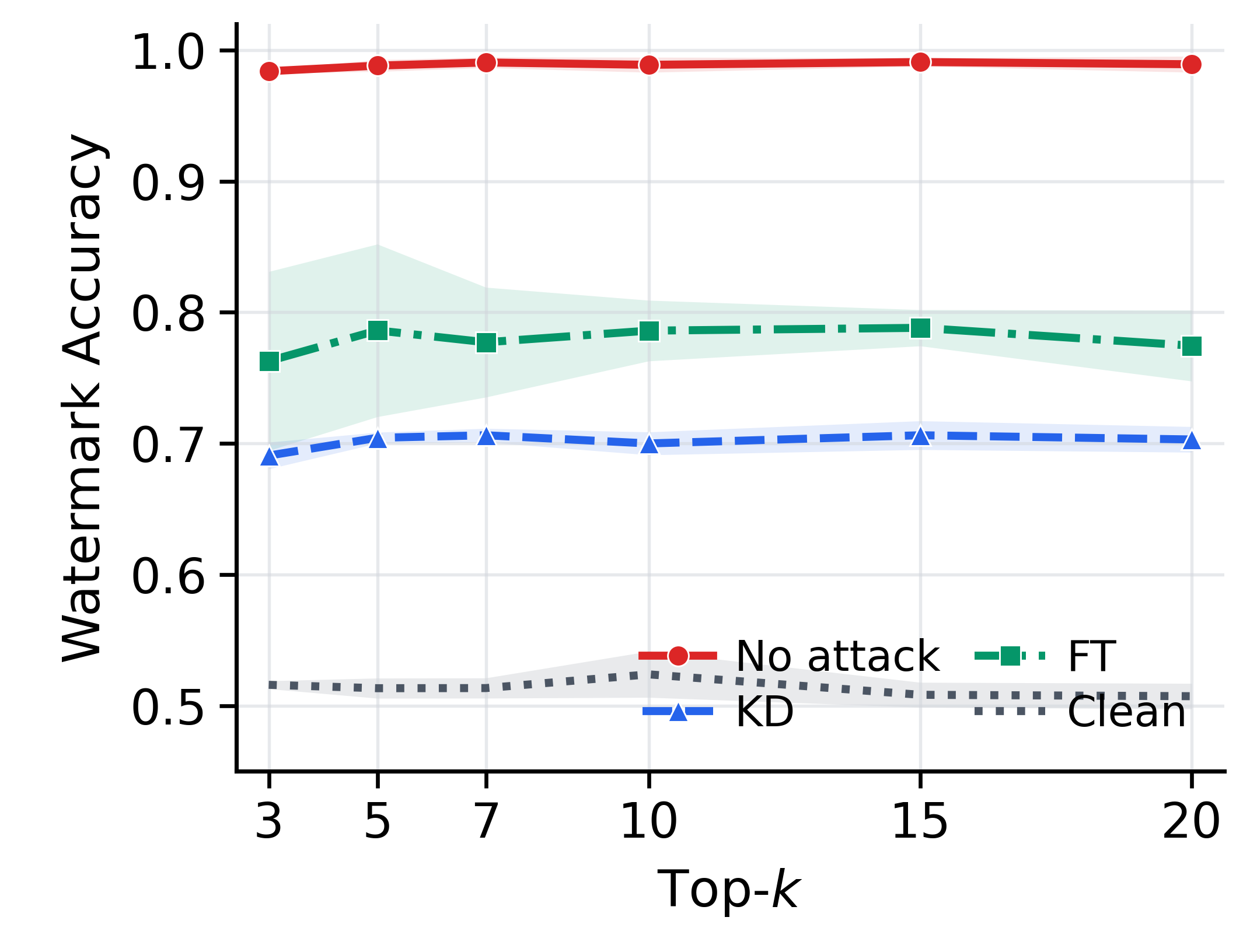}
    \caption{Robustness, 20News}
    \label{subfig:topk_rob_20news}
  \end{subfigure}
\caption{\textbf{Impact of Top-$k$.} Across CIFAR-100 and 20News, $k\geq3$ approaches full-output robustness while preserving main-task ACC, overcoming label-only capacity limits with negligible utility loss.\protect\footnotemark}
  \label{fig:topk_impact}
\end{figure}

\noindent\textbf{Top-3 Suffices for Robust Verification with Negligible Utility Loss.}
We empirically characterize the Pareto frontier (utility-robustness trade-offs) by varying the channel capacity $k$ from $k=1$ (label-only watermarks) to $k=K$ (full probability watermarks) on CIFAR-100 and 20News. In Figure~\ref{fig:topk_impact}, TOPMark exhibits a distinct \emph{utility-robustness decoupling}: increasing $k$ does not significantly affect main-task accuracy, and the gap between clean and watermarked models remains minimal. Watermark robustness, however, improves with larger $k$: Figure~\ref{subfig:topk_rob_cifar100} show greater resistance to KD and FT attacks as $k$ increases, while clean models remain near chance. Notably, $k \ge 3$ is sufficient to achieve strong utility and robustness, with only marginal gains beyond that; Figures~\ref{subfig:topk_fid_20news}--\ref{subfig:topk_rob_20news} show the same trend on 20News.

%This saturation has a practical implication: robust ownership evidence does not require exposing the full confidence vector. TOPMark retains a consistent verification rule as the API changes $k$, while limiting the information disclosed.

%main task Utility remains stable across the entire spectrum of $k$, confirming that embedding ownership signals into the Top-$k$ probability operates orthogonally to the semantic decision boundary. In contrast, robustness demonstrates a rapid asymptotic improvement, surging from the theoretical bottleneck at $k=1$ to near-perfect detection at $k \approx 10$ before saturating. This implies that full output exposure is unnecessary; a minimal expansion (e.g., $k=5$) is sufficient to reach the optimal Pareto efficient state, achieving maximum defense reliability with zero utility cost.

\subsection{Adaptive Attacks}

\footnotetext{Unlike Table~\ref{tab:TOPMark_performance}, this full $k=1,\ldots,K$ sweep does not exclude the top-3 logits, explaining the small ACC difference; see the Supplementary Material.}

\paragraph{Robustness against Logit-Layer Attacks.} We consider an API adversary that injects random noise $\xi$ with finite mean $\mu$ and variance $\sigma^2$ into output logits without knowing the private key $\mathcal{K}$. Assume that the per-logit noises are independent with the same finite variance. If the two available hash groups contain $n_0$ and $n_1$ logits, respectively, the variance of their mean-noise difference is $\sigma^2(1/n_0+1/n_1)$, while a common mean shift cancels in the comparative softmax. For balanced groups with $n_0=n_1=K/2$, each group mean has variance $2\sigma^2/K$ and their difference has variance $4\sigma^2/K$. Thus, larger available groups suppress independent noise, although the exact reduction depends on the observed top-$k$ set and any excluded logits. We test Gaussian noise applied to all logits, only non-maximal logits, or all except the top-3 or top-5 predictions. Supplementary Figure~4 shows that none substantially reduces ASR before causing severe utility loss; the stronger resilience on CIFAR-100 than CIFAR-10 is consistent with averaging over larger groups.

\paragraph{Robustness against More Black-Box Extraction Attacks.} We evaluate \emph{hard-label extraction}~\cite{tramer2016stealing}, \emph{data-free extraction}~\cite{Truong2021data}, and \emph{knowledge distillation} (KD)~\cite{Hinton_distilling_2014}, covering both label-only imitation and soft-output transfer. TOPMark retains 100\% TPR on CIFAR-10 despite ACC drops of 5.4--9.6 points; on CIFAR-100, data-free extraction and regularized KD also retain 100\% TPR. Hard-label extraction lowers TPR to 55\% only while reducing ACC from 75.6\% to 52.5\%, indicating that the attacker removes much of the task capability rather than selectively erasing the watermark. Full configurations and the complete utility--TPR results are reported in the Supplementary Material.

\begin{figure}[!t]
  \centering
  \begin{subfigure}{0.23\textwidth}
    \centering
    \includegraphics[width=\linewidth]{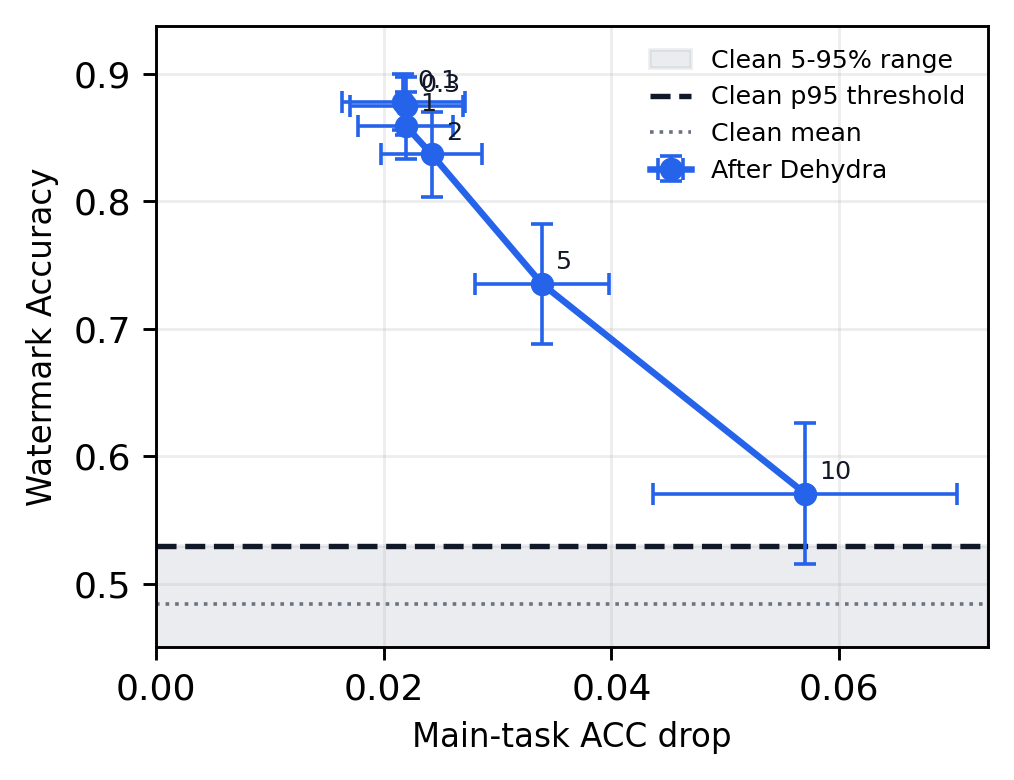}
    \caption{CIFAR-10}
    \label{subfig:dehydra_cifar10}
  \end{subfigure}
  \hfill
  \begin{subfigure}{0.23\textwidth}
    \centering
    \includegraphics[width=\linewidth]{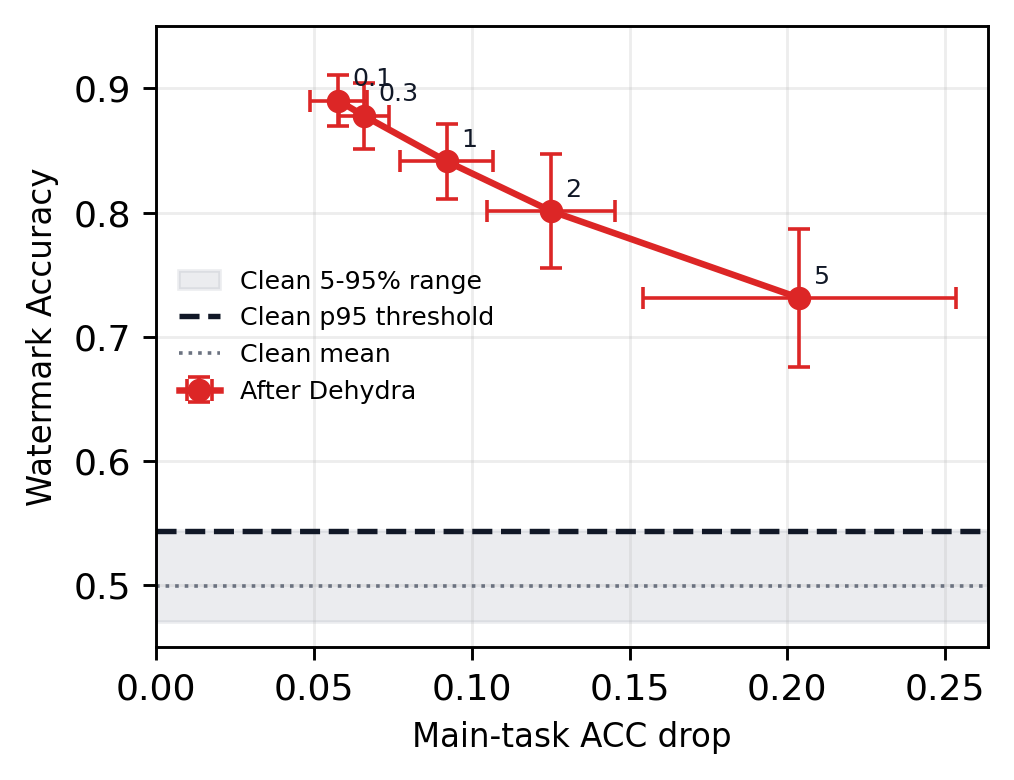}
    \caption{CIFAR-100}
    \label{subfig:dehydra_cifar100}
  \end{subfigure}
  \caption{\textbf{Neural Dehydration (Dehydra)~\cite{lu2024neural}.} Utility--watermark trade-off across attack strengths; error bars show run variation, and clean references denote the 5--95\% range, mean, and p95 threshold.}
  \label{fig:dehydra_tradeoff}
\end{figure}

\begin{figure}[!t]
  \centering
  \begin{subfigure}{0.23\textwidth}
    \centering
    \includegraphics[width=\linewidth]{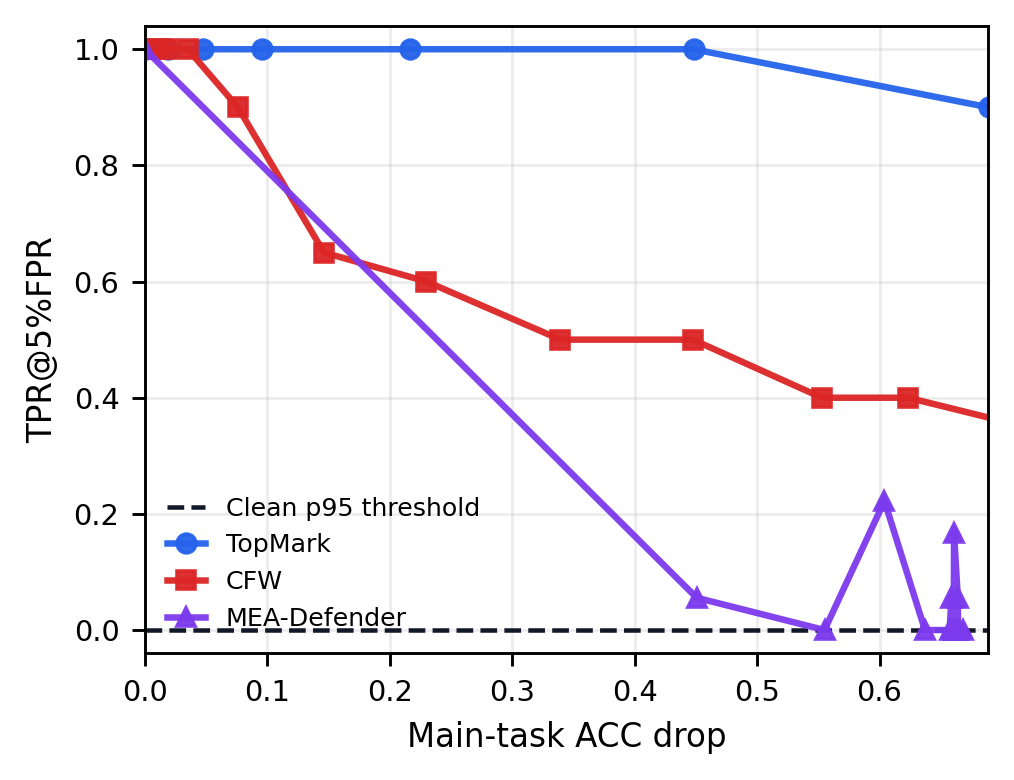}
    \caption{CIFAR-10}
    \label{subfig:advprune_cifar10}
  \end{subfigure}
  \hfill
  \begin{subfigure}{0.23\textwidth}
    \centering
    \includegraphics[width=\linewidth]{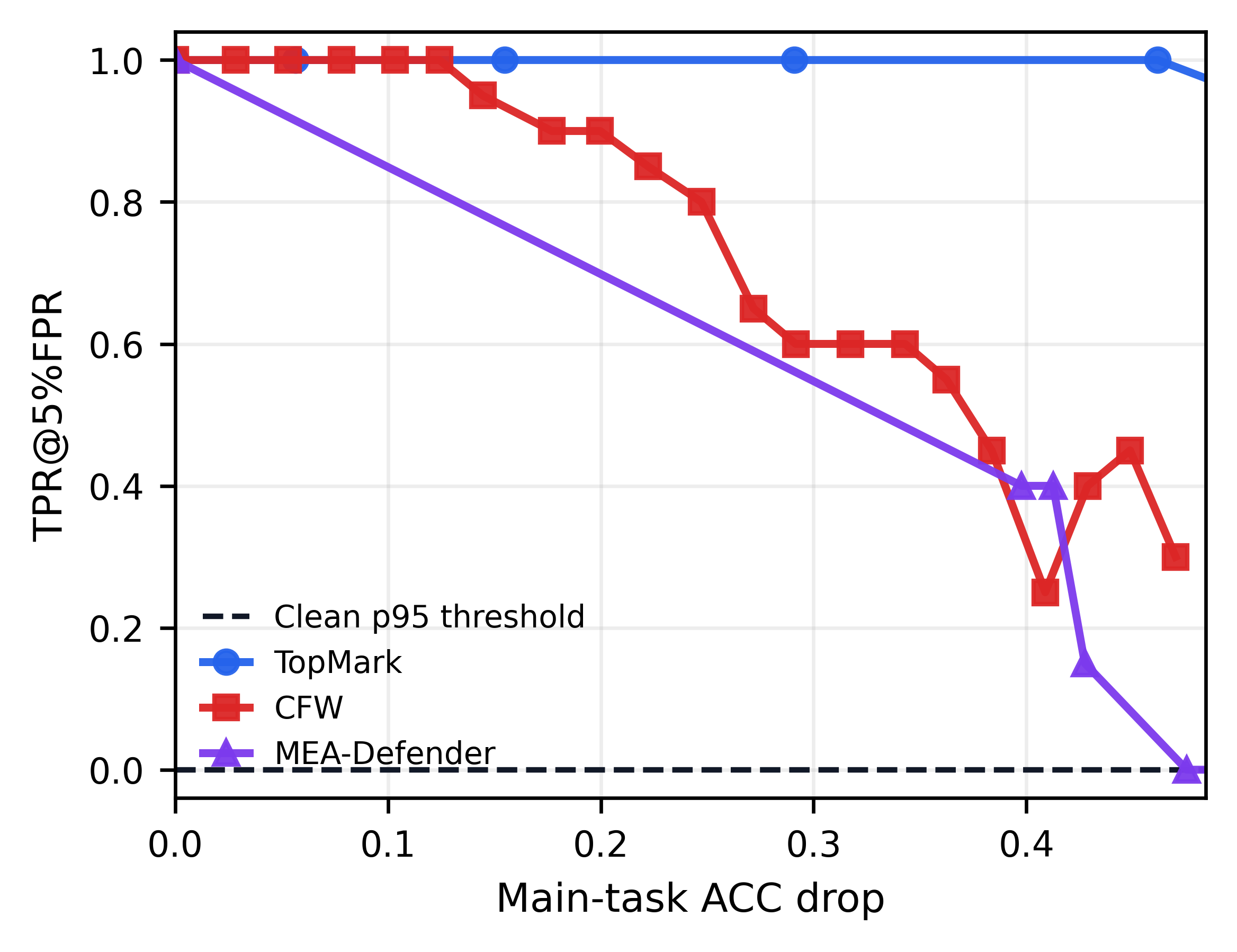}
    \caption{CIFAR-100}
    \label{subfig:advprune_cifar100}
  \end{subfigure}
  \caption{\textbf{Adaptive neuron pruning.} TPR@5\%FPR versus main-task ACC drop; TOPMark remains verifiable over a wider utility range than CFW and MEA-Defender.}
  \label{fig:adaptive_pruning}
\end{figure}

\paragraph{Robustness against Modern Watermark Erasure.} We evaluate Neural Dehydration (\textsc{Dehydra})~\cite{lu2024neural}, a watermark-agnostic \emph{white-box} attack with access to the victim's parameters and gradients. Without using the original watermark samples or trigger labels, Dehydra reconstructs class-conditioned synthetic samples from model internals and applies unlearning-style optimization to recover and remove latent watermark information while preserving task utility. Figure~\ref{fig:dehydra_tradeoff} shows that the mean post-attack score remains above the clean-model p95 threshold at every evaluated strength. Increasing the attack weight weakens the watermark only with greater main-task ACC loss: at the strongest CIFAR-10 setting, the uncertainty interval reaches the threshold region although its mean remains above it, while CIFAR-100 retains clearer separation. Thus, even with full white-box access, Dehydra does not reliably erase TOPMark within the evaluated utility range.

\section{Conclusion}
\label{sec:conclusion}

In this work, we revisit the conflict between predictive utility and watermark robustness in black-box ownership verification. Our information-theoretic analysis identifies the \emph{capacity crisis} of label-only verification: a single predicted label provides insufficient capacity to carry robust ownership evidence without interfering with semantic prediction. We introduce \textbf{TOPMark}, which uses top-$k$ probabilities, subspace binary classification, and a non-parametric cryptographic decoder to separate watermark verification from the task output. Across three domains, TOPMark achieves 99.5\% average TPR while preserving predictive utility. It remains verifiable under various removal attacks. 

\bibliography{ijcai26}

@inproceedings{liu2024falseclaims,
  author    = {Jian Liu and Rui Zhang and Sebastian Szyller and Kui Ren and N. Asokan},
  title     = {False Claims against Model Ownership Resolution},
  booktitle = {Proc. 33rd USENIX Secur. Symp.},
  pages     = {6885--6902},
  year      = {2024},
  publisher = {USENIX Association},
  url       = {https://www.usenix.org/conference/usenixsecurity24/presentation/liu-jian}
}

@article{liu2026attmark,
  author  = {Xinjing Liu and Zhuo Ma and Yang Liu and Taifeng Liu and Hao Yang and Zhan Qin},
  title   = {{AttMark}: Attention Based Model Watermarking Against Stealing Attacks},
  journal = {IEEE Trans. Dependable Secure Comput.},
  volume  = {23},
  number  = {1},
  pages   = {814--826},
  year    = {2026},
  doi     = {10.1109/TDSC.2025.3611654}
}

@article{zhao2026branchwm,
  author  = {Na Zhao and Kejiang Chen and Weiming Zhang and Nenghai Yu},
  title   = {Performance-Lossless Black-Box Model Watermarking},
  journal = {IEEE Trans. Dependable Secure Comput.},
  volume  = {23},
  number  = {2},
  pages   = {1955--1970},
  year    = {2026},
  doi     = {10.1109/TDSC.2025.3622253}
}

@article{li2025move,
  author  = {Yiming Li and Linghui Zhu and Xiaojun Jia and Yang Bai and Yong Jiang and Shu-Tao Xia and Xiaochun Cao and Kui Ren},
  title   = {{MOVE}: Effective and Harmless Ownership Verification via Embedded External Features},
  journal = {IEEE Trans. Pattern Anal. Mach. Intell.},
  volume  = {47},
  number  = {6},
  pages   = {4734--4751},
  year    = {2025},
  doi     = {10.1109/TPAMI.2025.3546223}
}

@misc{google_vision_api,
  title        = {Cloud Vision API: Label Detection},
  author       = {{Google Cloud}},
  year         = {{n.d.}},
  howpublished = {\url{https://cloud.google.com/vision/docs/labels}},
  note         = {Accessed 29 July 2026}
}

@inproceedings{barber2003algorithm,
  title={The {IM} Algorithm: A Variational Approach to Information Maximization},
  author={Barber, David and Agakov, Felix},
  booktitle={Advances in Neural Information Processing Systems (NIPS)},
  volume={16},
  pages={201--208},
  year={2003}
}

@inproceedings{fernandez2024functional,
  title={Functional invariants to watermark large transformers},
  author={Fernandez, Pierre and Couairon, Guillaume and Furon, Teddy and Douze, Matthijs},
  booktitle={ICASSP 2024-2024 IEEE International Conference on Acoustics, Speech and Signal Processing (ICASSP)},
  pages={4815--4819},
  year={2024},
  organization={IEEE}
}

@inproceedings{shamshadfirst,
  title={First-Place Solution to NeurIPS 2024 Invisible Watermark Removal Challenge},
  author={Shamshad, Fahad and Bakr, Tameem and Shaaban, Yahia Salaheldin and Hussein, Noor Hazim and Nandakumar, Karthik and Lukas, Nils},
  booktitle={The 1st Workshop on GenAI Watermarking at ICLR},
  year={2025},
  url={https://openreview.net/forum?id=wLaP37BrhE}
}

@article{sadasivan2025signature,
  title={Signature vs. Substance: Evaluating the Balance of Adversarial Resistance and Linguistic Quality in Watermarking Large Language Models},
  author={Guo, William and Uchendu, Adaku and Smith, Ana},
  journal={arXiv preprint arXiv:2511.13722},
  year={2025},
  doi={10.48550/arXiv.2511.13722}
}

@inproceedings{LiCLDZL23,
  author       = {Peixuan Li and
                  Pengzhou Cheng and
                  Fangqi Li and
                  Wei Du and
                  Haodong Zhao and
                  Gongshen Liu},
  editor       = {Brian Williams and
                  Yiling Chen and
                  Jennifer Neville},
  title        = {PLMmark: {A} Secure and Robust Black-Box Watermarking Framework for
                  Pre-trained Language Models},
  booktitle    = {Thirty-Seventh {AAAI} Conference on Artificial Intelligence},
  pages        = {14991--14999},
  publisher    = {{AAAI} Press},
  year         = {2023},
  url          = {https://doi.org/10.1609/aaai.v37i12.26750},
  doi          = {10.1609/AAAI.V37I12.26750},
}

@article{bai2025provfl,
  title={ProVFL: Property Inference Attacks against Vertical Federated Learning},
  author={Bai, Li and Zhang, Xinwei and Zhang, Sen and Ye, Qingqing and Hu, Haibo},
  journal={IEEE Transactions on Information Forensics and Security},
  volume={20},
  pages={6529--6543},
  year={2025},
  publisher={IEEE},
  doi={10.1109/TIFS.2025.3581743}
}

@inproceedings{GunnZS25,
  author       = {Sam Gunn and
                  Xuandong Zhao and
                  Dawn Song},
  title        = {An Undetectable Watermark for Generative Image Models},
  booktitle    = {The Thirteenth International Conference on Learning Representations,
                  {ICLR} 2025, Singapore, April 24-28, 2025},
  publisher    = {OpenReview.net},
  year         = {2025},

}

@inproceedings{Gloaguen0SV25,
  author       = {Thibaud Gloaguen and
                  Nikola Jovanovic and
                  Robin Staab and
                  Martin T. Vechev},
  title        = {Black-Box Detection of Language Model Watermarks},
  booktitle    = {The Thirteenth International Conference on Learning Representations,
                  {ICLR} 2025, Singapore, April 24-28, 2025},
  publisher    = {OpenReview.net},
  year         = {2025},
  url          = {https://openreview.net/forum?id=E4LAVLXAHW},
  bibsource    = {dblp computer science bibliography, https://dblp.org}
}

@article{ZhangCLMFZFHY24,
  author       = {Jie Zhang and
                  Dongdong Chen and
                  Jing Liao and
                  Zehua Ma and
                  Han Fang and
                  Weiming Zhang and
                  Huamin Feng and
                  Gang Hua and
                  Nenghai Yu},
  title        = {Robust Model Watermarking for Image Processing Networks via Structure
                  Consistency},
  journal      = {{IEEE} Trans. Pattern Anal. Mach. Intell.},
  volume       = {46},
  number       = {10},
  pages        = {6985--6992},
  year         = {2024},
  url          = {https://doi.org/10.1109/TPAMI.2024.3381543},
  doi          = {10.1109/TPAMI.2024.3381543},
  bibsource    = {dblp computer science bibliography, https://dblp.org}
}

@inproceedings{Shuo2025Explanation,
  author       = {Shuo Shao and
                  Yiming Li and
                  Hongwei Yao and
                  Yiling He and
                  Zhan Qin and
                  Kui Ren},
  title        = {Explanation as a Watermark: Towards Harmless and Multi-bit Model Ownership
                  Verification via Watermarking Feature Attribution},
  booktitle    = {32nd Annual Network and Distributed System Security Symposium, {NDSS}
                  2025, San Diego, California, USA, February 24-28, 2025},
  publisher    = {The Internet Society},
  year         = {2025},
  url          = {https://www.ndss-symposium.org/ndss-paper/explanation-as-a-watermark-towards-harmless-and-multi-bit-model-ownership-verification-via-watermarking-feature-attribution/},
  bibsource    = {dblp computer science bibliography, https://dblp.org}
}

@inproceedings{lu2024neural,
  title={Neural Dehydration: Effective Erasure of Black-box Watermarks from DNNs with Limited Data},
  author={Lu, Yifan and Li, Wenxuan and Zhang, Mi and Pan, Xudong and Yang, Min},
  booktitle={Proceedings of the 2024 on ACM SIGSAC Conference on Computer and Communications Security},
  pages={675--689},
  year={2024},
  doi={10.1145/3658644.3690334}
}

@misc{aws_rekognition_api,
  title        = {Amazon Rekognition DetectLabels API},
  author       = {{Amazon Web Services}},
  year         = {{n.d.}},
  howpublished = {\url{https://docs.aws.amazon.com/rekognition/latest/APIReference/API_DetectLabels.html}},
  note         = {Accessed 29 July 2026}
}

@article{guan2024world,
  title={World models for autonomous driving: An initial survey},
  author={Guan, Yanchen and Liao, Haicheng and Li, Zhenning and Hu, Jia and Yuan, Runze and Zhang, Guohui and Xu, Chengzhong},
  journal={IEEE Trans. Intell. Veh.},
  year={2024},
pages   = {1--17},
  publisher={IEEE}
}

@inproceedings{steinke2023privacy,
  author    = {Thomas Steinke and Milad Nasr and Matthew Jagielski},
  title     = {Privacy Auditing with One (1) Training Run},
  booktitle = {Proc. Adv. Neural Inf. Process. Syst. (NeurIPS)},
  volume    = {36},
  pages     = {49268--49280},
  year      = {2023}
}

@article{bonner2021current,
  title={Current best practice for presenting probabilities in patient decision aids: fundamental principles},
  author={Bonner, Carissa and Trevena, Lyndal J and Gaissmaier, Wolfgang and Han, Paul KJ and Okan, Yasmina and Ozanne, Elissa and Peters, Ellen and Timmermans, Dani{\"e}lle and Zikmund-Fisher, Brian J},
  title   = {Current Best Practice for Presenting Probabilities in Patient Decision Aids: Fundamental Principles},
  journal = {Med. Decis. Making},
  volume  = {41},
  number  = {7},
  pages   = {821--833},
  year    = {2021}
}

@ARTICLE{fan2022DeepIPR,
  author={Fan, Lixin and Ng, Kam Woh and Chan, Chee Seng and Yang, Qiang},
  journal={IEEE Transactions on Pattern Analysis and Machine Intelligence}, 
  title={DeepIPR: Deep Neural Network Ownership Verification With Passports}, 
  year={2022},
  volume={44},
  number={10},
  pages={6122-6139},
  doi={10.1109/TPAMI.2021.3088846}}

@inproceedings{zhu2024reliable,
  author    = {H. Zhu and S. Liang and W. Hu and F. Li and J. Jia and S.-L. Wang},
  title     = {Reliable Model Watermarking: Defending Against Theft without Compromising on Evasion},
  booktitle = {Proc. 32nd ACM Int. Conf. Multimedia (ACM MM)},
  pages     = {10124--10133},
  year      = {2024}
}

@article{Hinton_distilling_2014,
author = {Hinton, Geoffrey and Dean, Jeff and Vinyals, Oriol},
year = {2015},
month = mar,
journal={arXiv preprint arXiv:1503.02531},
pages = {1--9},
title = {Distilling the Knowledge in a Neural Network},
doi = {10.48550/arXiv.1503.02531}
}

@inproceedings{lukas2022sok,
  author    = {N. Lukas and E. Jiang and X. Li and F. Kerschbaum},
  title     = {SoK: How Robust Is Image Classification Deep Neural Network Watermarking?},
  booktitle = {Proc. IEEE Symp. Secur. Privacy (SP)},
  pages     = {787--804},
  year      = {2022}
}

@inproceedings{Lv2024MEA,
  author    = {Peizhuo Lv and Hualong Ma and Kai Chen and Jiachen Zhou and Shengzhi Zhang and Ruigang Liang and Shenchen Zhu and Pan Li and Yingjun Zhang},
  title     = {{MEA-Defender: A Robust Watermark against Model Extraction Attack}},
  booktitle = {Proc. IEEE Symp. Secur. Privacy (SP)},
  year      = {2024},
  pages     = {2515--2533},
  month     = may
}

@inproceedings{bansal2022certified,
  title={Certified Neural Network Watermarks with Randomized Smoothing},
  author={A. Bansal and P. Chiang and M. J. Curry and R. Jain and C. Wigington and V. Manjunatha and J. P. Dickerson and T. Goldstein},
  booktitle={Proc. Int. Conf. Mach. Learn. (ICML)},
  pages={1450--1465},
  year={2022},
   
}

@inproceedings{zhang2018protecting,
  title={Protecting Intellectual Property of Deep Neural Networks with Watermarking},
  author={J. Zhang and Z. Gu and J. Jang and H. Wu and M. P. Stoecklin and H. Huang and I. Molloy},
  booktitle={Proc. 2018 Asia Conf. Comput. Commun. Secur.},
  pages={159--172},
  year={2018},
  location={Incheon, Republic of Korea},
  month={May}
}

@article{yang_effectiveness_2019,
	title = {Effectiveness of Distillation Attack and Countermeasure on Neural Network Watermarking},
	journal = {arXiv},
	author = {Yang, Ziqi and Dang, Hung and Chang, Ee-Chien},
year = 2019,
month = jun
}

@inproceedings{jia2021entangled,
  title={Entangled Watermarks as a Defense against Model Extraction},
  author={H. Jia and C. A. Choquette-Choo and V. Chandrasekaran and N. Papernot},
  booktitle={Proc. 30th USENIX Secur. Symp. (USENIX Secur. 21)},
  pages={1937--1954},
  year={2021},
  month={Aug},
  isbn={978-1-939133-24-3}
}

@inproceedings{adi2018turning,
  title={Turning Your Weakness Into a Strength: Watermarking Deep Neural Networks by Backdooring},
  author={Y. Adi and C. Baum and M. Cisse and J. Keshet and B. Pinkas},
  booktitle={Proc. USENIX Secur. Symp.},
  pages={1615--1631},
  year={2018},
  month={Aug},
  location={Baltimore, MD, USA}
}

@inproceedings{rouhani2018deepsigns,
  title={DeepSigns: A Generic Watermarking Framework for IP Protection of Deep Learning Models}, 
  author={B. D. Rouhani and H. Chen and F. Koushanfar},
  booktitle={Proc. 24th Int. Conf. Archit. Support Program. Lang. Oper. Syst. (ASPLOS)},
  pages={485--497},
  year={2019},
  month={Apr}
}

@inproceedings{chen2021refit,
  title={{REFIT}: A Unified Watermark Removal Framework for Deep Learning Systems with Limited Data},
  author={X. Chen and W. Wang and C. Bender and Y. Ding and R. Jia and B. Li and D. Song},
  booktitle={Proc. 2021 ACM Asia Conf. Comput. Commun. Secur.},
  pages={321--335},
  year={2021},
  month={May}
}

@inproceedings{liu2018fine,
  title={Fine-Pruning: Defending Against Backdooring Attacks on Deep Neural Networks},
  author={K. Liu and B. Dolan-Gavitt and S. Garg},
  booktitle={Proc. Res. Attacks, Intrusions, Defenses (RAID)},
  pages={273--294},
  year={2018}
}

@article{wu2021adversarial,
  title={Adversarial Neuron Pruning Purifies Backdoored Deep Models},
  author={D. Wu and Y. Wang},
  journal={Adv. Neural Inf. Process. Syst.},
  volume={34},
  pages={16913--16925},
  year={2021}
}

@inproceedings{orekondy_knockoff_2019,
  author    = {T. Orekondy and B. Schiele and M. Fritz},
  title     = {Knockoff Nets: Stealing Functionality of Black-Box Models},
  booktitle = {Proc. IEEE/CVF Conf. Comput. Vis. Pattern Recognit. (CVPR)},
  pages     = {4949--4958},
  year      = {2019},
  month     = jun
}

@inproceedings{shokri2017membership,
	title={Membership inference attacks against machine learning models},
	author={Shokri, Reza and Stronati, Marco and Song, Congzheng and Shmatikov, Vitaly},
	booktitle={2017 IEEE Symposium on Security and Privacy (SP)},
	pages={3--18},
	year={2017},
	organization={IEEE}
}

@inproceedings{tramer2016stealing,
  author    = {F. Tram{\`e}r and F. Zhang and A. Juels and M. K. Reiter and T. Ristenpart},
  title     = {Stealing Machine Learning Models via Prediction APIs},
  booktitle = {Proc. 25th USENIX Secur. Symp.},
  pages     = {601--618},
  year      = {2016}
}

@inproceedings{yan2023rethinking,
  title={Rethinking White-Box Watermarks on Deep Learning Models under Neural Structural Obfuscation},
  author={Yan, Yifan and Pan, Xudong and Zhang, Mi and Yang, Min},
  booktitle={32th USENIX security symposium},
  year={2023}
}

@misc{hills_anadkat_shyamal_2023,
  author       = {Hills, James and Anadkat, Shyamal},
  title        = {Using logprobs},
  howpublished = {OpenAI Cookbook},
  month        = dec,
  year         = {2023},
  url          = {https://cookbook.openai.com/examples/using_logprobs},
  note         = {Accessed 29 July 2026}
}

@inproceedings{carlini2022membership,
  author    = {N. Carlini and S. Chien and M. Nasr and S. Song and A. Terzis and F. Tram{\`e}r},
  title     = {Membership Inference Attacks from First Principles},
  booktitle = {Proc. IEEE Symp. Secur. Privacy (SP)},
  pages     = {1897--1914},
  year      = {2022}
}

@inproceedings{cen2024transparency,
  author    = {S. H. Cen and R. Alur},
  title     = {From Transparency to Accountability and Back: A Discussion of Access and Evidence in {AI} Auditing},
  booktitle = {Proc. 4th ACM Conf. Equity Access Algorithms Mech. Optim. (EAAMO)},
  pages     = {1--14},
  year      = {2024}
}

@inproceedings{xiao2025class,
  title={Class-feature Watermark: A Resilient Black-box Watermark Against Model Extraction Attacks},
  author={Xiao, Yaxin and Ye, Qingqing and Liang, Zi and Li, Haoyang and Li, RongHua and Zheng, Huadi and Hu, Haibo},
  booktitle={Proceedings of the AAAI Conference on Artificial Intelligence},
  volume={40},
  pages={35903--35912},
  year={2026},
  doi={10.1609/aaai.v40i42.40905}
}

@inproceedings{Truong2021data,
  author       = {Jean{-}Baptiste Truong and
                  Pratyush Maini and
                  Robert J. Walls and
                  Nicolas Papernot},
  title        = {Data-Free Model Extraction},
  booktitle    = {{IEEE} Conference on Computer Vision and Pattern Recognition, {CVPR}
                  2021, virtual, June 19-25, 2021},
  pages        = {4771--4780},
  publisher    = {Computer Vision Foundation / {IEEE}},
  year         = {2021},
}

@inproceedings{charette2022coswm,
  author    = {Charette, Laurent and Chu, Lingyang and Chen, Yizhou and Pei, Jian and Wang, Lanjun and Zhang, Yong},
  title     = {Cosine Model Watermarking against Ensemble Distillation},
  booktitle = {Proceedings of the AAAI Conference on Artificial Intelligence},
  volume    = {36},
  pages     = {9512--9520},
  year      = {2022},
  doi       = {10.1609/aaai.v36i9.21184}
}

\end{document}